\documentclass[floats,twocolumn,prl,aps,superscriptaddress,unsortedaddress]{revtex4}
\usepackage{amssymb,amsmath,epsfig}

\def\msun{M_{\odot}}
\def\lya{Ly$\alpha~$}

\begin{document}

\title{Baryon Acoustic Oscillation Intensity Mapping as a Test of  Dark Energy}


\author{Tzu-Ching Chang}\email{tchang@cita.utoronto.ca}
\affiliation{Inst. for
  Astronomy and Astrophysics, Academia Sinica, PO Box 23-141, Taipei
  10617, Taiwan}\affiliation{CITA, University of Toronto, 60
  St.George St., Toronto, ON, M5S 3H8, Canada }

\author{Ue-Li Pen}\email{pen@cita.utoronto.ca}
\affiliation{CITA, University of Toronto,
  60 St.George St., Toronto, ON, M5S 3H8, Canada
} 
\author{Jeffrey B. Peterson}\email{jbp@cmu.edu}
\affiliation{Department of Physics, Carnegie Mellon University,
  500 Forbes Ave, Pittsburgh, PA 15213, USA
}
\author{Patrick McDonald}\email{pmcdonal@cita.utoronto.ca}
\affiliation{CITA, University of Toronto,
  60 St.George St., Toronto, ON, M5S 3H8, Canada
}

\begin{abstract}
The expansion of the universe appears to be accelerating, and the
mysterious anti-gravity agent of this acceleration has been called
``dark energy''.  To measure the dynamics of dark energy, Baryon
Acoustic Oscillations (BAO) can be used. Previous discussions of the
BAO dark energy test have focused on direct measurements of redshifts
of as many as $10^9$ individual galaxies, by observing the 21cm line
or by detecting optical emission.  Here we show how the study of
acoustic oscillation in the 21 cm brightness can be accomplished by
economical three dimensional intensity mapping.  If our estimates
gain acceptance they may be the starting point for a new class of dark
energy experiments dedicated to large angular scale mapping of the
radio sky, shedding light on dark energy.
\end{abstract}

\maketitle

\def\citep#1{\cite{#1}}

{\it Introduction.---} To understand dark energy and sharply test
theories of its character, it is necessary to precisely measure the
last half of the expansion history. This period corresponds to the
redshift range $0 < z \lesssim 2$. Many techniques have been proposed
to study the late-stage expansion history, and some of the most
promising make use of Baryon Acoustic Oscillations
(BAO)\cite{Blake:2003}. In addition to producing CMB structure, the
acoustic oscillations also produced density structure in the atomic
gas and dark matter, which is still detectable today.  Many groups
have reported 2 to 3 $\sigma$ detections, in the low redshift
universe, of periodic structures in the density of galaxies at the
predicted wavelengths across the sky\cite{Percival:2007}.  Because the
acoustic waves are frozen in after recombination, the BAO peak
wavelengths can be used as a cosmological standard ruler: observation
of the angular size of the peak wavelengths across a range of
redshifts allows accurate measurement of the expansion history.

In this letter we present calculations of structure in the three
dimensional brightness due to the hyperfine transition of neutral
hydrogen (HI) at 21 cm wavelength.  We show that, via 21 cm emission,
baryon oscillations could be precisely measured, using a telescope
just 200 wavelengths in diameter, since each cosmic cell of the
appropriate scale contains more than $10^{12}\msun$ of emitting
neutral hydrogen gas.  We present forecasts on the dark energy
constraints from this new type of observation, which are competitive
with the best proposed dark energy experiments\cite{detf:2006}.  Where
needed in the analysis we adopt {\it WMAP3} values for the
cosmological parameters\citep{Spergel:2007}, $\Omega_m=0.24,
\Omega_b=0.04$, $\Omega_\Lambda=0.76$, and $h=0.73$, where $\Omega_m$,
$\Omega_b$, and $\Omega_\Lambda$ are the matter, baryon, and dark
energy fractional density, respectively, and $h$ is the dimensionless
Hubble parameter. We also use the {\it WMAP3} error limits for these
values.

A recent paper proposed a test of dark energy models by measurement of
baryon oscillations in the 21 cm brightness field at redshifts $z >3$
\cite{Wyithe:2007}.  Using only such high redshift data two
cosmological parameters---departures of spatial flatness $\Omega - 1$
and a slow change of dark energy equation of state---are nearly
indistinguishable. It is only when precise flatness $\Omega =1$ is
assumed that such high redshift data can be used to constrain dark
energy models. In this letter we use the more standard approach and
assume that $\Omega$ is not perfectly measured. Then data at lower
redshift become essential if dark energy models are to be tightly
constrained.

{\it Detectability.---} To measure the oscillations it is not
necessary to detect individual galaxies, but only to measure the
variation in HI mass on large scales.  A familiar analogy is the study
of galaxies via images of their surface brightness.  Except for the
nearest of galaxies, we do not detect the individual stars that
produce the optical emission, indeed the pixels of most galaxy images
represent the emission of millions of stars.  Never-the-less such
images allow study of galactic structure.

Above the third BAO peak nonlinear evolution attenuates BAO structure,
so the third peak is at the smallest spatial scale we need
consider. This peak has a wavelength 35$h^{-1}$ Mpc. A Nyquist sampled
map therefore needs pixels of size 18$h^{-1}$ Mpc. At $z=1.5$ the
corresponding angular wavelength is 20 arc minutes, which requires a
telescope of approximately 200 wavelength, or 100 meters, to resolve.

We base our estimate of neutral hydrogen mass contained in a
$18h^{-1}$ Mpc volume ( $M_{\rm HI} \sim 2 \times 10^{12} \msun$) on \lya
absorption studies of the cosmic density of neutral gas
\cite{Wolfe:2005}, which find $\Omega_{\rm HI} \sim 1 \times 10^{-3}$
at $z\sim 1$, with a weak dependence on redshift.

The average sky brightness temperature due to the 21cm line is about
$300 \mu$K.  The mean brightness temperature $T_b$ of the 21cm line
can be estimated using\cite{Barkana:2007}
\begin{equation}
T_b = 0.3\left({\Omega_{\rm HI} \over 10^{-3}}\right)
\left({\Omega_m + a^{3} \Omega_\Lambda\over 0.29}\right)^{-1/2}  
\left({1+z \over 2.5}\right)^{1/2} {\rm mK}.
\label{eq:21cm}
\end{equation}
Most observation of the sky are measurements of variations in temperature $\Delta T_b \sim T_b \delta$,   
where $1+\delta = \rho_g / {\bar \rho_g}$ is the normalized neutral
gas density, and $a=(1+z)^{-1}$ is the scale factor.

\begin{figure}[t]
\centerline{\epsfig{file=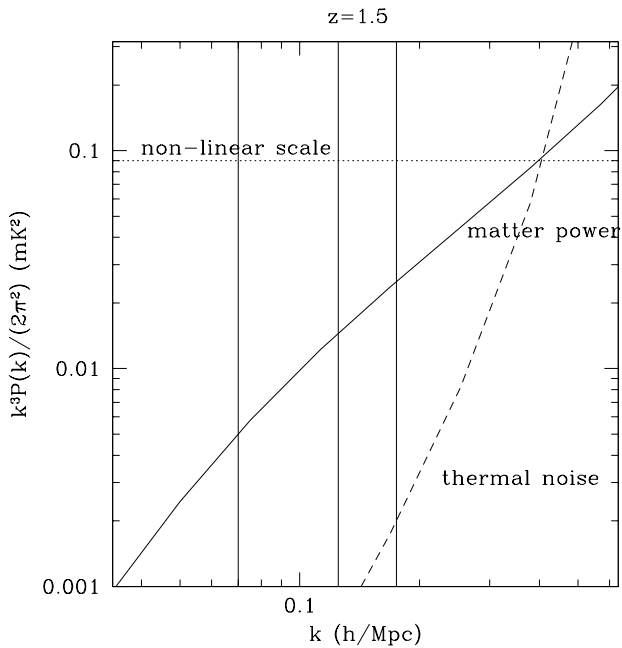, width=\columnwidth}}
\caption{The power spectrum of neutral hydrogen hyperfine emission at
  $z=1.5$ is shown by the solid line.  The dashed line indicates the
  map noise from a 200-by-200-meter radio cylinder
  telescope operated in transit mode for 100 days, assuming a
  telescope system temperature of 50 K.  The horizontal dotted line
  indicates the non-linear scale at $z=1.5$. The vertical lines
  indicate the location of peaks in Figure~\ref{fig:pk_bao}
\label{fig:hshs}}
\end{figure}

\begin{figure}[t]
\centerline{\epsfig{file=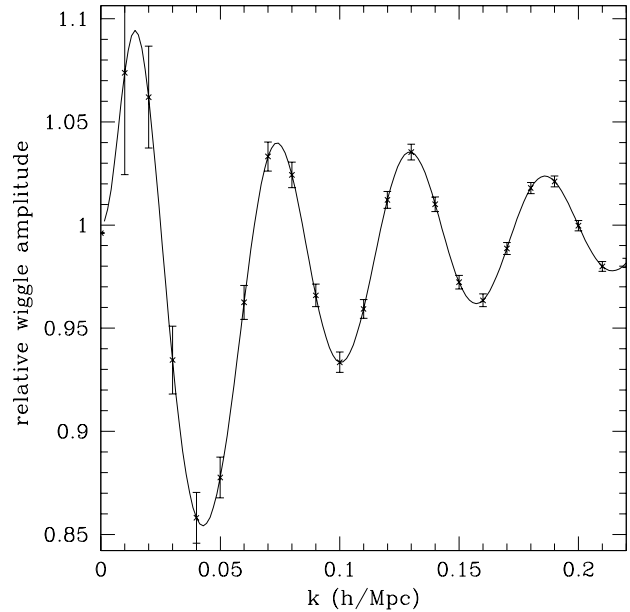, width=\columnwidth}}
\caption{Baryon Acoustic Oscillations. To show
  BAO  we plot the ratio of matter power spectrum that
  contains contributions from baryons, calculated using a fitting
  formula\cite{Eisenstein:1998}, over a power spectrum with no
  baryons. To single out the oscillation, we further divide the
  ratio by a smooth fitting curve to the overall spectrum.  The error
  bars represent projections of the sensitivity possible with a
  200-by-200-meter cylinder telescope.}
\label{fig:pk_bao}
\end{figure}

On 18$h^{-1}$ Mpc scales spatial variations of the sky brightness due
to 21 cm emission in the cosmic web structure presents a sky noise at
the 150 $\mu$K level.  Figure~\ref{fig:hshs} shows the amplitude of
this cosmological large scale structure, while Figure~\ref{fig:pk_bao}
plots the small oscillations imprinted by BAO.  The variation of gas
density due to baryon oscillations is small compared to the
fluctuation of the large-scale structure.  To detect the BAO signals
beneath this noise Fourier analysis of a large regions of sky is
required.  The observations should cover as much sky as possible.

For a filled aperture telescope the RMS variation in measured sky
temperature $\Delta T =T_{\rm sys} / \sqrt{\Delta f ~t_{\rm int}}$,
where $\Delta T$ is the temperature in Kelvin, $T_{\rm sys}$ is the
system temperature of the radio telescope in Kelvin, $\Delta f$ is the
observing bandwidth in Hertz and $t_{int}$ the integration time in
seconds.  Assuming current technology, which allows $T_{\rm sys} =50
$K, one needs an observing bandwidth-time product of $2 \times
10^{11}$ to bring the thermal map noise down to $\sim 100 \mu$K, below
the cosmic structure noise.  In frequency space 18 $h^{-1}$ Mpc spans
roughly 3 MHz at $z=1.5$, so the required observation time is $\sim
6\times 10^4$ seconds, about 18 hours per pixel.  Small regions of sky
might be mapped at this noise level using existing telescopes such as
the Green Bank Telescope (GBT), the Effelsberg 100-meter Radio
Telescope or the Parkes Telescope.

Brightness sensitivity is not improved by increase of telescope
aperture, instead multiple receivers, either on a single reflector or
on individual reflectors, can be used to increase the accumulated
integration time.

For sensitivity estimates, we consider an intensity mapping
telescope, dubbed IM, covering a square
aperture of size 200m $\times$ 200m.  We subdivide this area into 16
cylindrical sections, each 12.5m wide and 200m long, analogous to the
Hubble Sphere Hydrogen Survey (HSHS)\cite{Peterson:2006}.  The
cylinders are aligned North-South allowing the telescope to
instantaneously sample almost the entire meridian.  This geometry allows
all-sky coverage each 24 hour period in a telescope with no moving
parts. 
The estimated sensitivity of the survey is shown in Figure~\ref{fig:hshs}.

{\it Foregrounds.---} The desired BAO signal is smaller than other
sources of sky brightness structure.  The largest interference is due
to synchrotron emission from the Galaxy. The spatial structure of this
emission exceeds the BAO signal by $10^3$.  Similar emission by
extragalactic radio sources distributed across the sky will contribute
sky brightness fluctuations on smaller scales.  However all these
foregrounds are spectrally smooth, with brightness temperature $T
\propto f^\alpha$, so one expects to be able to subtract them
spectrally. The image from each 3 MHz slice of the frequency spectrum
contains a unique BAO pattern, largely uncorrelated with the
neighboring slices.  In contrast, the foregrounds and backgrounds are
highly correlated. By subtracting the average of both neighboring
slices the BAO signal can be retained while the other emission is
removed.  Of course this subtractions will be imperfect, because of
imperfections in the telescope and spatial variations in the
astrophysical source spectrum. A similar challenge is faced by the
higher-redshift 21cm experiments probing the Cosmic Reionization era
and many foreground subtraction methods have been proposed
(\cite{Furlanetto:2006} and references therein).

\begin{figure}
\centerline{\epsfig{file=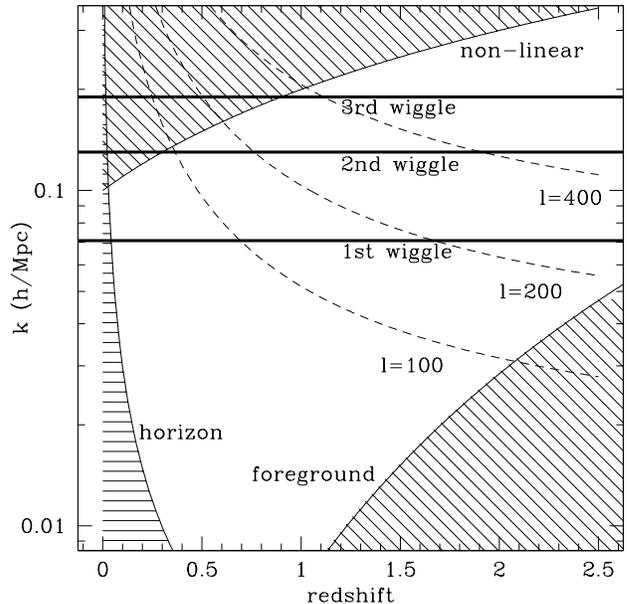, width=\columnwidth}}
\caption{The observable parameter space in redshift and in scale (k)
  for BAO. The shaded regions are observationally inaccessible (see
  text). The horizontal lines indicate the scale of the first three
  BAO wiggles, and the dashed lines show contours of constant spherical harmonic order $\ell$.} 
\label{fig:fg}
\end{figure}

The spectral index $\alpha$ for Galactic emission varies across the
sky.  The frequency-double-difference technique described above
suppresses the Galactic sky brightness but leaves a residual of order
$\Delta T = \Delta \alpha (\Delta f / f)^{2} T$ to first power in
$\Delta \alpha$, where $\Delta \alpha$ is the uncertainty in spectral
index and $\Delta f$ the frequency difference.  Using the {\it WMAP3}
measurement we estimate

\begin{equation} 
\Delta T \sim 1.3 ~\Delta \alpha \left({l \over
  200}\right)^{-0.5} \left({1+z \over 2.5}\right)^{3.2} \left({\Delta
  f \over f} \right)^2  ~~{\rm K},
\label{eq:fg}
\end{equation}
where $l$ is the spherical harmonic number.   The normalization factor
of 1.3 Kelvin temperature fluctuation is derived by extrapolating from
the {\it WMAP3} measurement of Galactic foreground variation at 20 GHz
on 3.7$^{\circ}$ degree scale, along with the spectral index of
synchrotron emission at mid-Galactic latitude\cite{Kogut:2007}.
Adopting $\Delta\alpha=1$, a conservative upper limit, we estimate a
residual Galactic interference signal which is plotted in Figure~\ref{fig:fg}.

\begin{table*}
\begin{ruledtabular}
\begin{tabular}{lcccccccccccccc}
$ $ & $\Omega_m h^2$ & $\Omega_b h^2$ & $\Omega_{\rm DE}$ & $a_p$ & $w_p$ &
 $w'$ & $\Omega_k$ & $\log_{10}(A)$ & $n_s$ & $h$ & $\Omega_m$ &
${\rm FoM}$ \\
${\rm reference\  value}$ & $0.128$ & $0.0223$ & $0.760$ & $$ & $-1.00$ & 
$0.00$ & $0.00$ & $-8.64$ & $0.950$ & $0.730$ & $0.240$ & $$ \\
P+IM & $0.0012$ & $0.00017$ & $0.0058$ & $0.65$ & $0.015$ & $0.16$ & $0.00072$ & $0.0077$ & $0.0062$ & $0.0075$ & $0.0054$ & $421.9$ \\ 
P+IMW & $0.0012$ & $0.00017$ & $0.010$ & $0.63$ & $0.023$ & $0.29$ & $0.00084$ & $0.0077$ & $0.0062$ & $0.015$ & $0.010$ & $152.3$ \\ 
P+II & $0.0012$ & $0.00017$ & $0.012$ & $0.79$ & $0.036$ & 
$0.52$ & $0.0031$ & $0.0076$ & $0.0061$ & $0.019$ & $0.012$ &$53.39$
\\
P+II+III & $0.0011$ & $0.00015$ & $0.0045$ & $0.76$ & $0.020$ 
& $0.18$ & $0.0021$ & $0.0063$ & $0.0046$ & $0.0070$ & $0.0043$ &
$279.5$ \\
P+II+III+IV & $0.00091$ & $0.00013$ & $0.0028$ & $0.75$ & $0.011$ &
$0.090$ & $0.0013$ & $0.0050$ & $0.0028$ & $0.0045$ &
$0.0026$ & $975.2$ \\
P+II+III+IV+IM & $0.00068$ & $0.00013$ & $0.0025$ & $0.69$ & $0.0082$ & $0.068$ & $0.00047$ & $0.0049$ & $0.0028$ & $0.0032$ & $0.0023$ & $1793.$ \\ 
P+II+III+IV+IMW & $0.00071$ & $0.00013$ & $0.0027$ & $0.71$ & $0.0095$ & $0.073$ & $0.00063$ & $0.0050$ & $0.0028$ & $0.0036$ & $0.0025$ & $1440.$ \\
\end{tabular}
\end{ruledtabular}
\caption{
Fiducial cosmological parameters and error bars for combining the Planck mission with the fiducial BAO
intensity mapping experiment such as IM (P+IM), with the Dark Energy Task Force\cite{detf:2006} (DETF) Stage II projects (P+II), plus the Stage III projects (P+II+III), plus the Stage IV projects (P+II+III+IV), and all above combined (P+II+III+IV+IM).  Here IM is the conservative best guess for intensity mapping experiment parameters, while IMW is the worse case scenario (see text).  DETF Stage II is well-defined which includes weak lensing (WL), cluster evolution (CL), and supernova (SN) experiments; for Stage III we use the optimistic photometric versions of WL, CL, and SN; for Stage IV we use the optimistic space-based WL, CL, and SN experiments.  The table
entries are shown using the units of the DETF Report: IM compares favorably
with other proposed Stage III projects.  The dark energy equation of
state is parametrized as $w=w_p+(a_p-a)w'$.  $\Omega_{\rm DE}$ and $\Omega_k$ are the dark energy and curvature density, respectively.  
$n_s$ is the spectral
index of the primordial power spectrum, and $A$ its normalization.
The Figure of Merit (FoM) refers to the inverse volume in the 
95\% confidence ellipse.
\label{tab:par}}
\end{table*}

Figure~\ref{fig:fg} 
shows various potential limitations to BAO
measurements using 21 cm sky brightness mapping.  The top exclusion 
comes from
the non-linear scale, which erases the baryon wiggles.  The left
exclusion comes from the finite volume, which sets a lower bound on
the lowest wavenumber that fits within a given volume.  The lower
right is a rough estimate of the impact of foregrounds using the
double frequency difference described above. The shaded region
reflects the region where the estimated residual foreground from
Eqn. (\ref{eq:fg}) exceeds the expected signal in
Eqn. (\ref{eq:21cm}).


{\it Parameter Estimation.---} To estimate the sensitivity of this
technique for measurement of dark energy parameters, we follow the
procedure in Seo and Eisenstein (2007) \cite{Seo:2007} and carry out
the standard figure-of-merit calculation used in the DETF
report\cite{detf:2006}.  Our calculations include the effects of
redshift space distortion and a simplified model of non-linearity
using a smoothing window.  This type of analysis, the conversion from
survey data to cosmological parameters, is an area of active research,
and several effects such as mode-mode coupling are still being
investigated.  Our use of shells of 21 cm surface brightness
introduces two new types of unknowns into the analysis: the HI
density, and the bias of 21 cm sources.  To accommodate these
uncertainties we make one sensitivity estimate using conservative
values for the unknown quantities (IM) as our baseline model, and one
with a worst case assumption that each of these effects works to
increase the error (IMW). The IM case estimate assumes that
$\Omega_{\rm HI}$ remains at the present day value of 0.0005,
independent of redshift, and we adopt a bias $b=1$.  The shot noise
from present day HI galaxies \cite{Zwaan:2005} is negligible.
Motivated by the reconstruction of BAO degradation due to
non-linearity in \cite{Eisenstein:2007} the Seo and Eisenstein damping
scale is shortened by a factor of two.  For the worst case (IMW)
estimate we let $\Omega_{\rm HI}$ to be 0.0004.  We adopt a bias
$b=0.8$, which is the current estimate of the low $z$ value. This
increases the noise, and seems securely to be a worst case because
galaxies are unlikely to be both antibiased and rare.  We assume that
21 cm sources are rare and massive at high redshift, and we use the
Lyman-Break Galaxy luminosity function (with a density of 0.004
h$^3$ Mpc$^{-3}$) for the shot noise calculation.  The Seo and Eisenstein
damping scale is assumed.

\begin{figure}
\centerline{\epsfig{file=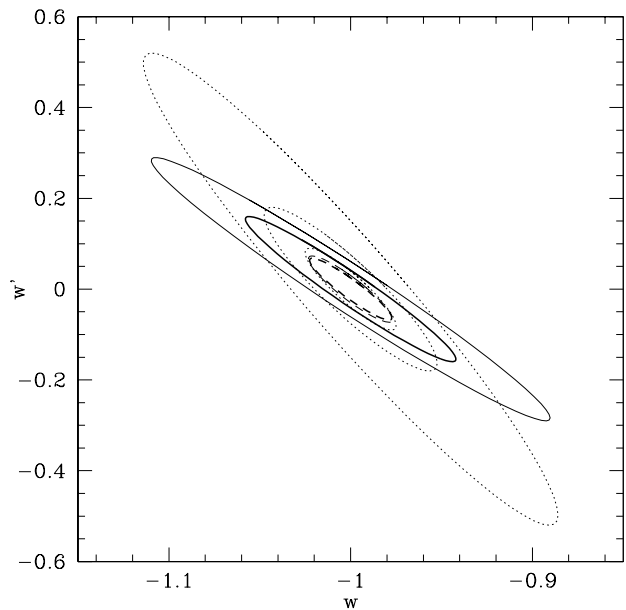, width=\columnwidth}}
\caption{The 1-$\sigma$ contour for IM combined with Planck (inner thick solid for
baseline model, outer thin solid for worst case), the Dark Energy Task Force Stage II projects with Planck (outer dotted), the Stage II and III projects with Planck (intermediate dotted), the Stage II, III and IV projects with Planck (inner dotted), and all above experiments combined (dashed, again thick for baseline, thin for worst case;  the two contours are nearly indistinguishable).
} 
\label{fig:planck}
\end{figure}

The time variable dark energy density is
parametrized as $\rho_{DE} \propto (1+z)^{3(1+w)}$; an ordinary
cosmological constant corresponds to $w=-1$, and the principal goal of
dark energy experiments is to measure deviations of $w$ from $-1$.  We
first determine the Fisher matrix for the tangential and radial
distances errors.  From these, we compute the errors on dark energy
parameters.  The resulting projected errors to the equation of state
are shown in Figure~\ref{fig:planck}.  Table \ref{tab:par} quantifies
the errors, which can be compared to other dark energy surveys.

The mapping of cosmological large scale structure by three dimensional 21 cm intensity mapping
is an approach that can be tested with existing  
telescopes today, which appears to offer very competitive constraints on dark energy parameters.
This new tool may soon allow astronomers to economically map huge volumes of the
universe.  

{\it Acknowledgments.---}
We acknowledge financial support by NSERC.

\bibliographystyle{apsrev}
\bibliography{bao}
\end{document}